\RequirePackage{fix-cm}
\documentclass[twocolumn]{svjour3}          
\smartqed  
\usepackage{graphicx}

\usepackage{amsfonts}

\usepackage{amsmath}

\usepackage{amssymb}

\usepackage{longtable}
\usepackage{bm}
\usepackage{multirow}
\usepackage{hyperref}
\usepackage{color}
\hypersetup{ colorlinks,
linkcolor=blue,
filecolor=black,
urlcolor=blue,
citecolor=blue }

\usepackage{subfigure}

\usepackage{sidecap}

 \usepackage{mathptmx}      
\begin{document}

\title{Mechanical strain in capped and uncapped self-assembled Ge/Si quantum dots}

\author{P. F. Gomes \and  H. A. Fernandes        \and         J. L. Gonz\'{a}lez-Arango
}

\institute{P. F. Gomes,  H. A. Fernandes  \at
              Instituto de Ci\^encias Exatas e Tecnol\'{o}gicas, Universidade Federal de Goi\'{a}s, CEP 75801-615, Jata{\'i}, Brazil.
              \email{paulofreitasgomes@ufg.br}  
           \and
           J. L. Gonz\'{a}lez-Arango \at
              Departamento De Fisica y Geologia, Universidad de Pamplona, Pamplona, Colombia.
}

\date{Received: date / Accepted: date}

\maketitle

\begin{abstract}
In this study we numerically calculate the spatial profile of mechanical strain on self-assembled germanium (Ge) quantum dots (QDs) grown on a silicon (Si) substrate. Although the topic has been exhaustively studied, interesting features have not been explained or even mentioned in the literature yet. We studied the effect of the cap layer considering two cases: capped QDs (where a Si cap is present above the Ge QDs) and uncapped QDs (where no Si is present above the Ge QDs). We observed that Ge in the capped QDs is more strained compared with the the uncapped QDs. This expected effect is attributed to the additional tension from the Si cap layer. However, the situation is opposite for the Si substrate, it is more strained in the uncapped QD because the Ge layer is less strained in this case. We also calculated the band-edge alignment for the electrons and holes. 
\end{abstract}

\section{Introduction}  \label{intro}

Mechanical strain plays an important role in electronic, optical, and transport properties of semiconductor quantum dots (QDs) \cite{stangl,boxberg2007,jacobsen2006}. Epitaxial dots have a spatial strain profile caused by the mismatch of lattice parameters of the involved materials. This strain creates an elastic energy which is one of the determining factors in the formation of dots in the Stranski-Krastanov \cite{ribeiro2007nanoletters} mode. The strain also changes the potential profile experienced by charger carriers, necessary to calculate the electronic eigenstates \cite{grundmann1995,andreev2000}. Several methods to calculate the strain profile are available, which have been successfully applied to semiconductor nanostructures. The most used methods involve the continuous elasticity \cite{jogai2001,cros2006} or atomistic models \cite{stier1999,he2004}. Several theoretical and experimental works on germanium (Ge)/silicon (Si) QDs can be found in the literature \cite{yakimov2010,brunner,kurdi2006,gomes2007}, although none of them describes in detail the features of the strain and tension in uncapped QDs.

In the present work, we studied the effect of the cap layer on the mechanical strain on self-assembled Ge QDs grown on silicon a Si substrate (Ge/Si QDs). We calculated the strain profile by numerically solving the equilibrium equations in the continuous elasticity model using finite element method. The rest of this paper is organized as follows: in the next section, we present the model and describe the parameters considered in the calculations. In Section 3, we show the results of our simulations. In Section 4, we present our conclusions.

\section{Theoretical model}  

\subsection{Linear elastic theory}

Axial symmetry is an approximation where a three-dimensional structure is generated by the revolution of a two-dimensional one. Dome (or lens-shaped) Ge QDs \cite{ross1999,gilbertoscience} belong to this case. This approximation requires that all physical parameters of a crystalline lattice possess this symmetry, which means that all the relevant quantities of the problem do not depend on the angular coordinate $\varphi$ of the cylindrical coordinates $(r,\varphi,z)$. This approximation allows the problem to be solved in the two-dimensional plane $r,z$ for $r>0$ and it is very useful because it considerably decreases the computational cost of the simulations.

The model used in this work is based on the numerical solution of the equilibrium equations of the linear elastic theory \cite{landau,sad2004}. This theory is suitable for treating materials with small strain and is based on the Hooke's law, namely, $\mathbf{F} = - k\mathbf{x}$. Using a two-order tensor for strain $e_{ij}$ and stress $\sigma_{ij}$, Hooke's law is expressed as \cite{montoro2010}:
\begin{equation}
\begin{bmatrix}
\sigma_x  \\
\sigma_y  \\
\sigma_z  \\
\sigma_{xy}  \\
\sigma_{yz}  \\
\sigma_{zx}  \\
\end{bmatrix}
=
\begin{bmatrix}
C_{11} & C_{12} & C_{12} &   0  &  0  &  0  \\
C_{12} & C_{11} & C_{12} &   0  &  0  &  0  \\
C_{12} & C_{12} & C_{11} &   0  &  0  &  0    \\
  0    &   0    &   0    & C_{44}     &  0  &  0 \\
  0    &   0    &   0    &   0   & C_{44}   &  0 \\
 0    &   0    &   0    &   0   &  0   & C_{44}  \\ 
\end{bmatrix}
\begin{bmatrix}
e_x  \\
e_y  \\
e_z  \\
e_{xy}  \\
e_{yz}  \\
e_{zx}  \\
\end{bmatrix}
\end{equation}
where $C_{ij}$ is the elastic constants matrix of a material with cubic symmetry. Conversely, if the material has a cylindrical symmetry, we need to perform a coordinate transformation (equivalent to $x \rightarrow r$ and $y \rightarrow \varphi$) to obtain the Hooke's law in cylindrical coordinates. In addition, by considering axial symmetry approximation, no quantity has angular dependence. Then, in cylindrical coordinates we have
\begin{equation}
\sigma_{r\varphi} = \sigma_{\varphi z} = e_{r\varphi} = e_{\varphi z} = \frac{dv}{d\varphi} = 0, \label{supondo}
\end{equation}
where $v$ is the displacement in the $y-$direction. Therefore, the Hooke's law can be simplified as
\begin{equation}
\begin{bmatrix}
\sigma_r  \\
\sigma_{\varphi}  \\
\sigma_z  \\
\sigma_{rz}  \\
\end{bmatrix}
=
\begin{bmatrix}
C_{11} & C_{12} & C_{12} &   0    \\
C_{12} & C_{11} & C_{12} &   0    \\
C_{12} & C_{12} & C_{11} &   0    \\
  0    &   0    &   0    & C_{44} \\
\end{bmatrix}
\begin{bmatrix}
e_r  \\
e_{\varphi}  \\
e_z  \\
e_{rz}  \\
\end{bmatrix}.
\label{leidehookesimetriaaxial}
\end{equation}

The stress tensor is defined in terms of the external force as $\vec{\nabla} \cdot \sigma_{ij} + \vec{F} = 0$. Because we do not consider any displacement in the center of the mass, we obtain $\vec{F} = 0$; therefore
\begin{equation}
\vec{\nabla} \cdot \sigma_{ij}= 0. \label{eqlkjdfassss}
\end{equation}
Substituting Eq. (\ref{leidehookesimetriaaxial}) into Eq. (\ref{eqlkjdfassss}) yields
\begin{equation}
\vec{\nabla} \cdot \left( C_{ij} \delta e_{ij} \right) =0. \label{defor-relaxacaocompacot}
\end{equation}
where the solution for the strain is represented as $\delta e_{ij}$. To obtain this solution, we numerically solve the above differential equation [Eq. (\ref{defor-relaxacaocompacot})] using the software Comsol Multiphysics.

\begin{figure}
\centering
\includegraphics[width=2.6 in]{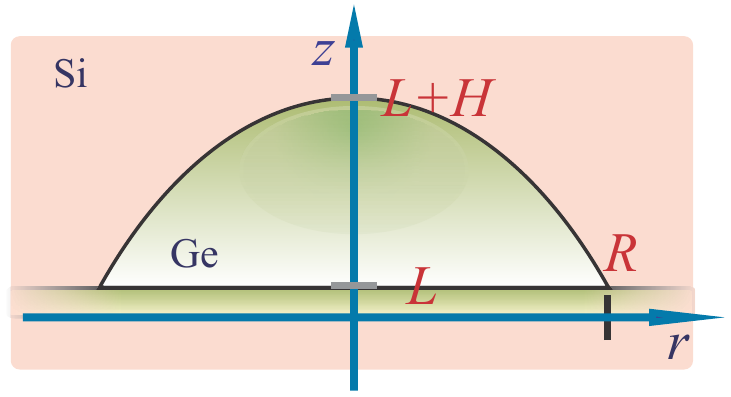} 
\caption{Illustration of the dome-shaped geometry modeled as a spherical cap. The dome is the top of a sphere whose radius is 43.8 nm.}
\label{geometrias}
\end{figure}

\subsection{Initial condition $\varepsilon$ for the strain}

The components of the total strain are expressed as follows:
\begin{eqnarray}
e_r &=& \chi \varepsilon_r + \delta e_r \nonumber \\
e_{\varphi} &=& \chi \varepsilon_{\varphi} + \delta e_{\varphi} \nonumber \\
e_z &=& \chi \varepsilon_z + \delta e_z \nonumber  \\
e_{rz} &=& \delta e_{rz} \nonumber 
\end{eqnarray} 
where $\varepsilon_i$ is the components of the initial value of the strain needed to solve Eq. (\ref{defor-relaxacaocompacot}). The characteristic function $\chi$ is defined as:
\begin{equation}
\chi(\mathbf{r}) = \left\{
\begin{array}{rl}
1 & \text{if  } \mathbf{r} \in \Omega_{\text{Ge}}  \\
0 & \text{if  } \mathbf{r} \in \Omega_{\text{Si}}
\end{array} 
\right.
\label{funcaocaracterstica22}
\end{equation}
where $\Omega_{\text{Ge}}$ and $\Omega_{\text{Si}}$ refer to the space domains occupied by Ge and Si, respectively. The purpose of $\chi$ is to take into account the assumption that the initial strain is zero in the Si domain and constant and nonzero in the Ge domain. In addition,  the shear components of the initial strain are set to zero, i.e.,
\begin{equation}
\varepsilon_{rz} = \varepsilon_{r\varphi} = \varepsilon_{\varphi z} = 0. \label{elskfjwwwwjl}
\end{equation}

\begin{figure}
\subfigure{a)
\includegraphics[width=3.2 in]{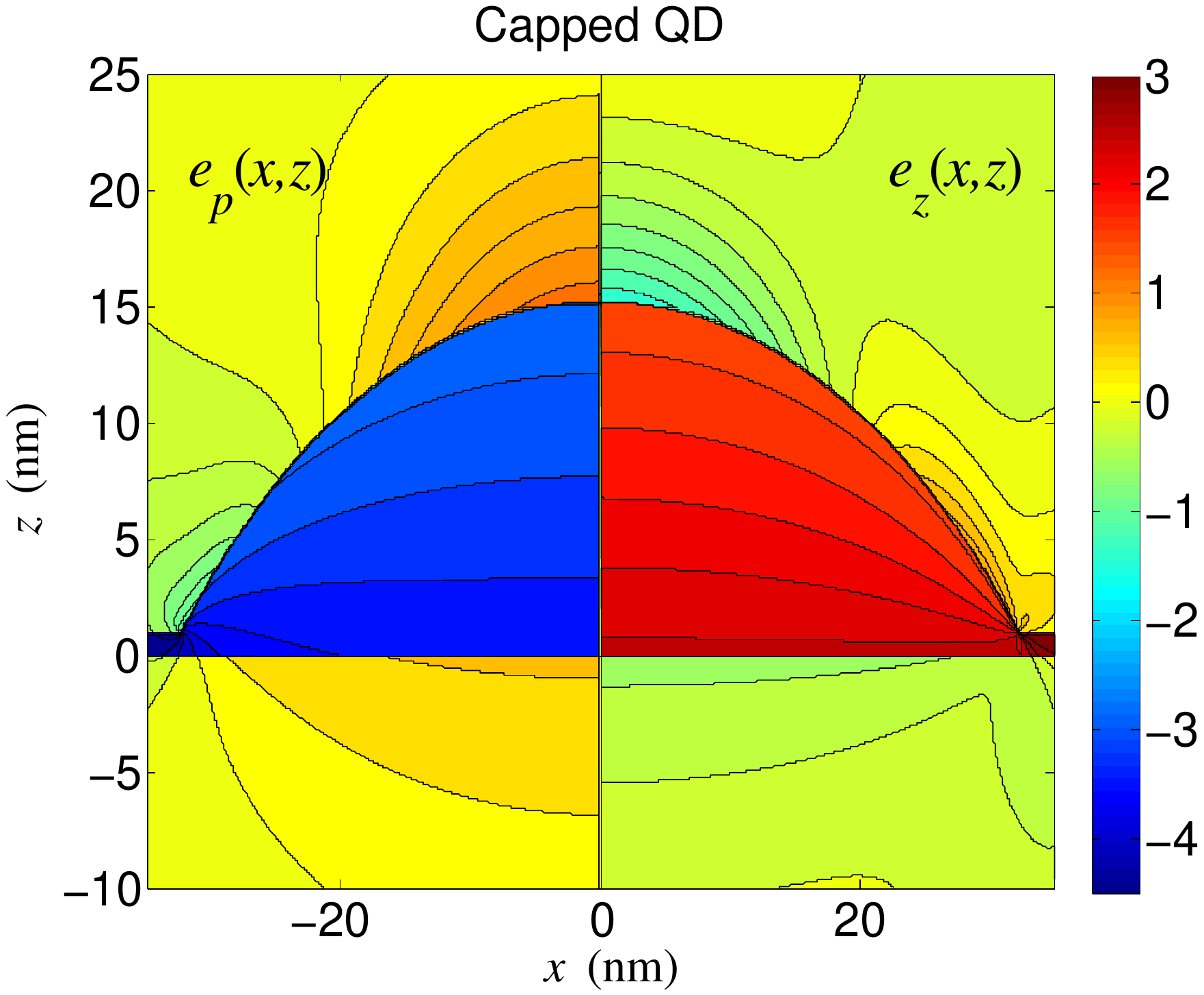}
\label{epzca}
}
\subfigure{b)
\includegraphics[width=3.2 in]{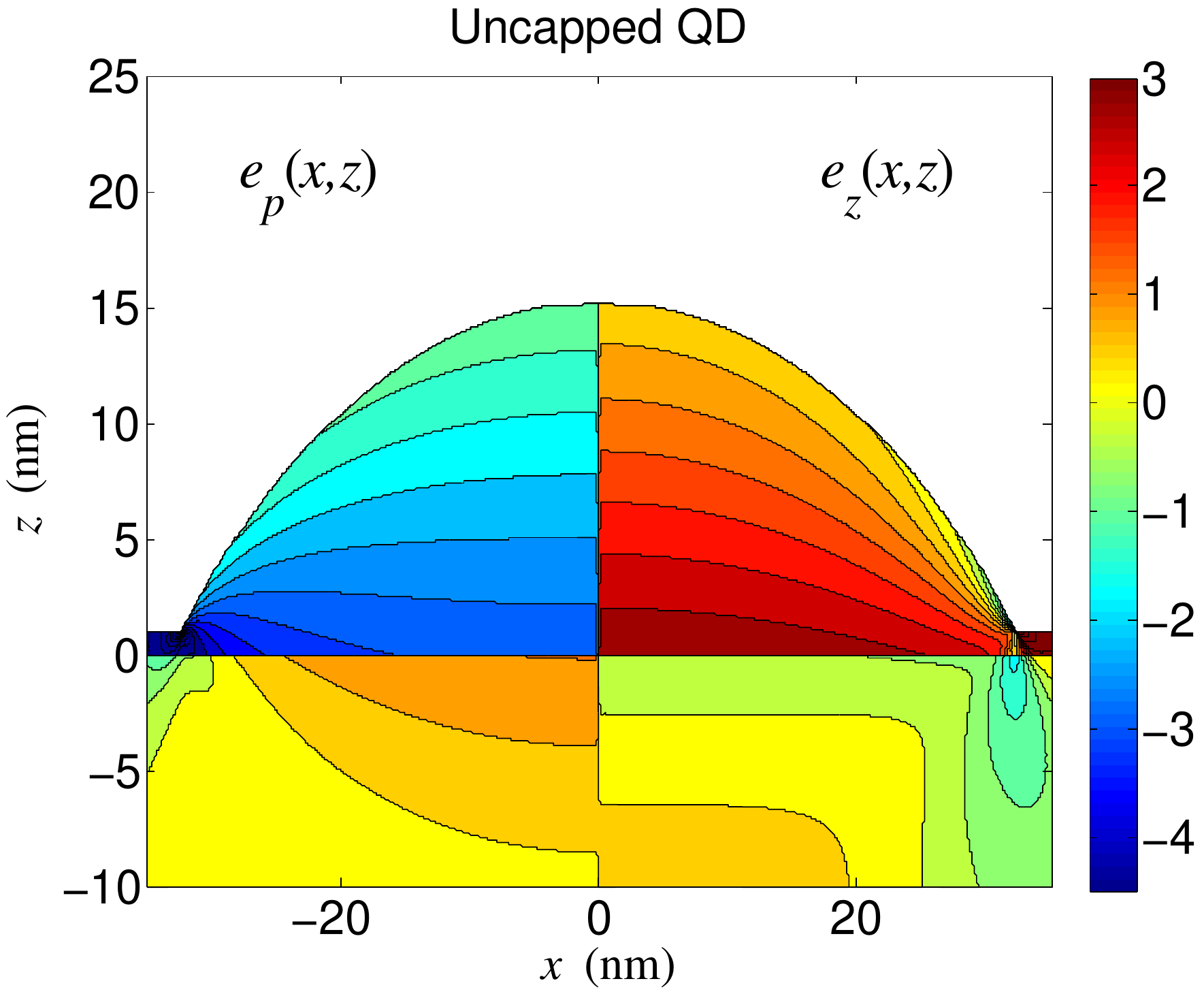}
\label{epzun}
}
\caption{Spatial profile of $e_p(r,z)$ [Eq. (\ref{deformacao-plano})] (on the left side) and $e_z(r,z)$ (on the right side) for \subref{epzca} Capped QD and \subref{epzun} Uncapped QD. The strain is multiplied by 100 to be expressed in \% and the color scale bar is the same for both graphics.}
\end{figure}

Hereafter, we show how to calculate the initial values. In an epitaxial film containing QDs (with lattice parameter $a_i$) grown over a substrate (001) (with parameter $a_m$), the strain in the plane is biaxial and given by the mismatch in lattice parameter $\alpha$, which is defined as \cite{kuo2008}
\begin{equation}
\alpha =\frac{a_m-a_i}{a_i}. \label{descasamentoparametrrede}
\end{equation}
Where we used the the following assumption:
\begin{equation}
\varepsilon_r =  \varepsilon_{\varphi} = \alpha. \label{deformacaobiaxialplanolkj}
\end{equation}
In this condition, the stress tensor can be approximated by: 
\begin{equation}  \label{tensao-inicial}
\sigma_r = \sigma_{\varphi} = 0. 
\end{equation}
Substituting Eqs. (\ref{deformacaobiaxialplanolkj}) and (\ref{tensao-inicial}) into the Hooke's law yields \cite{kuo2008}
\begin{equation} 
\varepsilon_z = -2 \alpha \frac{C_{12}}{C_{11}}. \label{epsilonzinifid}
\end{equation} 

\subsection{Lattice parameter and stress tensor}

The strain $e_z(r,z)$ gives the relative variation in the distance between nearest neighbors in the $z-$direction. In plane $xy$, the strain is the average of the other two components:
\begin{equation}
e_p(r,z) = \frac{1}{2} (e_r +e_{\varphi}). \label{deformacao-plano} 
\end{equation}

The lattice parameter in plane $xy$ is defined as \cite{jalabert}:
\begin{equation}
a_r(r,z) =  \left[1+e_{r}(r,z)\right] \left[\chi a_{\text{Ge}} +(1-\chi)a_{\text{Si}} \right], \label{presplano}
\end{equation}
where $a_{\text{Ge}}$ and $a_{\text{Si}}$ are the lattice parameters of Ge and Si, respectively, and $\chi$ is defined in Eq. (\ref{funcaocaracterstica22}). We also calculated the hydrostatic and biaxial components of the strain, which are defined as \cite{kuo2008}:
\begin{equation}
e_h (r,z)= e_z + e_p \label{hidrostatico} \\
\end{equation}
and
\begin{equation}
e_b (r,z)= e_z - e_p, \label{biaxial}
\end{equation}
respectively. The hydrostatic component reflects the variation in the volume of the structure due to the strain. From Hooke's law, the stress tensor is expressed as:
\begin{equation}
\begin{bmatrix}
\sigma_r  \\
\sigma_{\varphi}  \\
\sigma_z  \\
\sigma_{rz}  \\
\end{bmatrix}
=
\begin{bmatrix}
C_{11} & C_{12} & C_{12} &       0    \\
C_{12} & C_{11} & C_{12} &       0    \\
C_{12} & C_{12} & C_{11} &       0    \\
  0       &     0     &      0    &   C_{44} \\
\end{bmatrix}
\begin{bmatrix}
\chi \alpha  + \delta e_r(r,z) \\
\chi \alpha  + \delta e_{\varphi}(r,z) \\
\chi \varepsilon_z  + \delta e_z (r,z) \\
\delta e_{rz}(r,z) \\
\end{bmatrix}.
 \label{leidehookedetalhada}
\end{equation}

\subsection{Band-edge alignment}

The difference in gap energy of the materials (in our case: Ge and Si) creates a band offset acting as potential well for the carriers. We represent this potential by $U_c$ [for the electrons in the conduction band (CB)], and by $U_v$ [for the holes in the valence band (VB)]. The hydrostatic component $e_h$ shifts this potential and the biaxial one $e_b$ brakes the degeneracy of CB splitting the sixfold-degenerate $\Delta$ valleys into the fourfold $\Delta(4)$ and twofold $\Delta(2)$. In VB, the biaxial strain also splits the heavy and light hole bands. We write $\Phi_{\Delta 2}$ and $\Phi_{\Delta 4}$ for the energy levels of the electrons and $\Psi_{hh}$ and $\Psi_{lh}$ for the energy levels of the heavy and light holes \cite{gomes2007}. Their dependency with the strain are given by
\begin{eqnarray}
\Phi_{\Delta 2} (r,z,e_h,e_b) &=& U_c(r,z) + C_{\Delta 2}(e_h,e_b) \label{phidelta2} \\
\Phi_{\Delta 4} (r,z,e_h,e_b) &=& U_c(r,z) + C_{\Delta 4}(e_h,e_b) \label{phidelta4} \\ 
\Psi_{hh} (r,z,e_h,e_b) &=& U_v(r,z) + V_{hh}(e_h,e_b) \label{psihh} \\ 
\Psi_{lh} (r,z,e_h,e_b) &=& U_v(r,z) + V_{lh}(e_h,e_b) \label{psilh} 
\end{eqnarray}
The confinement potentials are defined as
\begin{eqnarray}
U_c (r,z) &=& E_{s} + \chi U_{BC}  \nonumber \\
U_v (r,z) &=& \chi U_{BV} \nonumber
\end{eqnarray}
where $E_{Si}=1.17$ eV is the energy gap of the Si at 0 K. The contributions of the strain in each potential are written as \cite{Pryor1998,Schmidt2000,Califano2002}:
\begin{eqnarray}
C_{\Delta 2}(e_h,e_b) &=& a_h e_h + \frac{2}{3} d  e_b \nonumber \\
C_{\Delta 4}(e_h,e_b) &=& a_h e_h - \frac{1}{3} d  e_b \nonumber \\
V_{hh}(e_h,e_b) &=& \dfrac{\Delta_0}{3} - be_b \nonumber \\
V_{lh}(e_h,e_b) &=& -\dfrac{\Delta_0}{6} + \frac{1}{2} be_b + \frac{1}{2} \sqrt{(\Delta_0 -9be_b)^2 +8 \Delta_0 b e_b}, \nonumber 
 \end{eqnarray}
 where $a_h = a_c-a_v$. The importance of these band-edge energies is that they enter as input in the Schroedinger equation in order to calculate the wave-functions of the carriers. Therefore, a minimum in the potential can represent a localization of the wave function.

\subsection{Calculation parameters}

With regard to the geometry of the QD, we relied on the results obtained by Magalhães-Paniago \textit{et. al} \cite{paniago}. They observed a dome-shaped geometry with average radius $R = 32.3$ nm and height $H = 14.2$ nm (Fig. \ref{geometrias}). We also considered a quantum well (also called wetting layer) with thickness $L = 1$ nm between $z=0$ and $z=L$ whereas the QD is located between $L < z < L+H$. Using nanometer (nm) as the units for space, the Si substrate comprises the domains $-500<z<500$ and $0<r<500$. The wetting layer comprises $0<r<500$ and $0<z<1$. The calculation was done only in the region $r>0$. From Eq. (\ref{descasamentoparametrrede}) and Table \ref{parametros} we have
\begin{equation}
\alpha = \dfrac{a_{Si}-a_{Ge}}{a_{Ge}} = -0.039. \nonumber
\end{equation}

\begin{table}
\centering
\begin{tabular}{|c|cc|}
\hline
Quantity & Si & Ge  \\
\hline
\hline $a$ (\AA) & 5.431  & 5.657   \\ 
\hline $C_{11}$ ($10^{11}$ Pa) & 1.66 & 1.29  \\ 
\hline $C_{12}$ ($10^{11}$ Pa) & 0.64 & 0.48  \\ 
\hline $C_{44}$ ($10^{11}$ Pa) & 0.8 & 0.67  \\ 
\hline  $a_c$ (eV) & -3.133 & -1,786 \\
\hline  $a_v$ (eV) & -3.1 & -4.54 \\
\hline  $d$ (eV) & 8.6 & 9.4 \\
\hline  $\Delta_0$ (eV) & 0.044 & 0.296 \\
\hline  $b$ (eV) & -2.1 & -2,86 \\
\hline
\end{tabular}
\caption{Constants and parameters of Si and Ge. The values were obtained from Ref. \cite{brunner}.}
\label{parametros}
\end{table}

\begin{figure}
\subfigure{a)
\includegraphics[width=3.0 in]{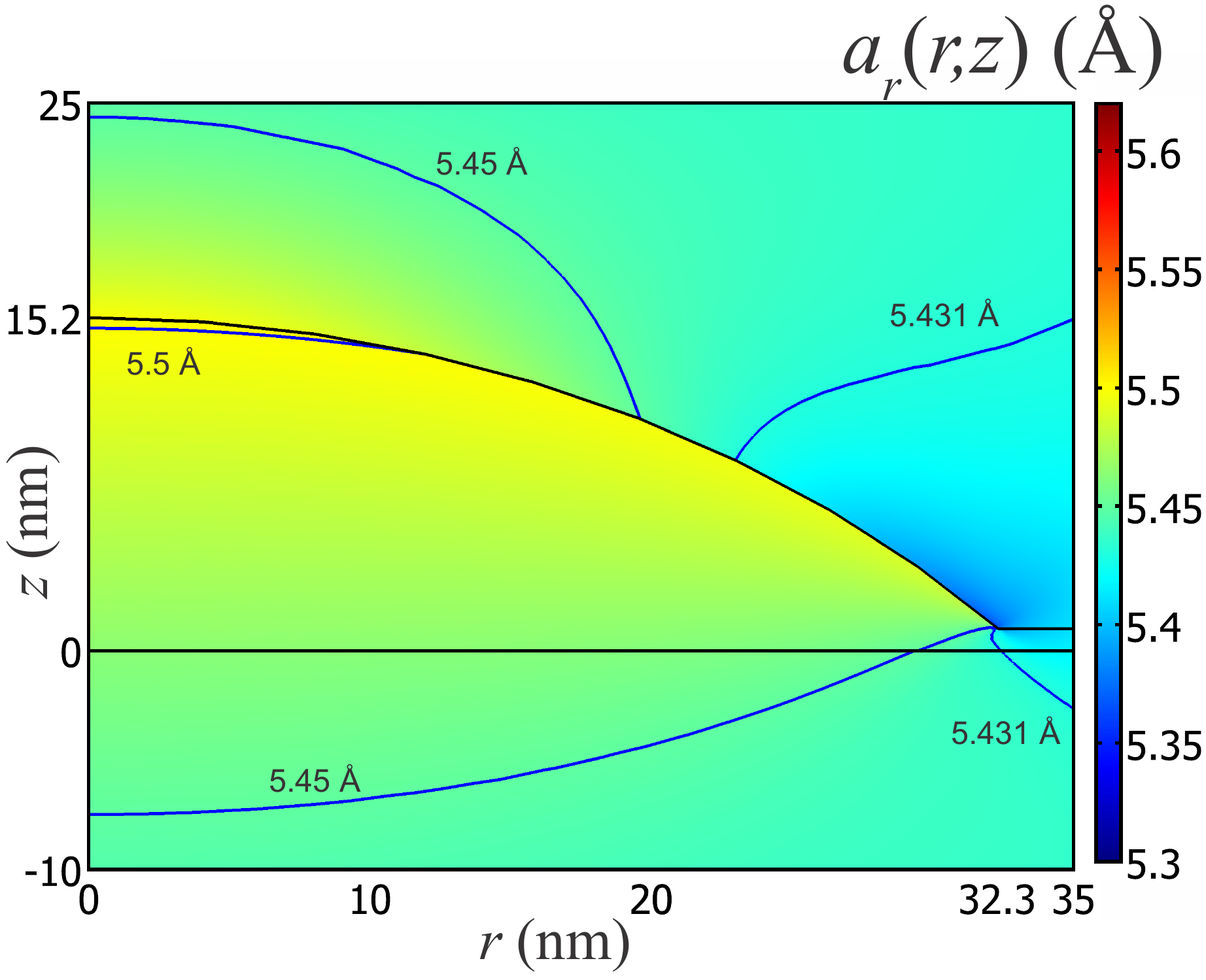}
\label{domo-ap-coberto}
}
\\
\subfigure{b)
\includegraphics[width=3.0 in]{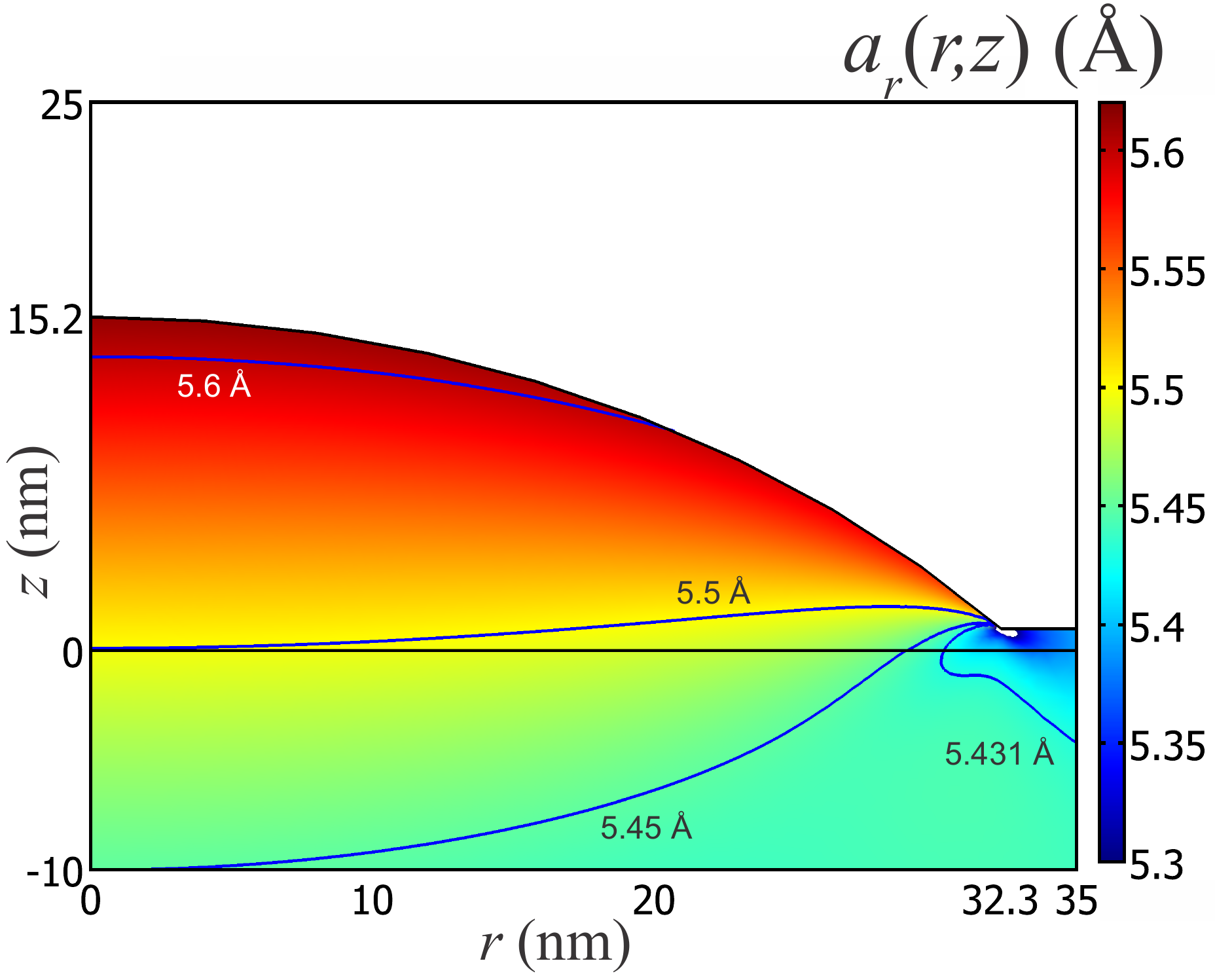}
\label{domo-ap-descoberto}
}
\caption{Spatial profile (in color) of the resultant lattice parameter in the plane $a_r (r,z)$ [Eq. (\ref{presplano})]. The blue lines indicate constant values of $a_r$. \subref{domo-ap-coberto} Capped QD. \subref{domo-ap-descoberto} Uncapped QD. The color scale bar is the same for both graphics.}
\end{figure}

\section{Results}

Figure \ref{epzca} shows the spatial profile of the strain in plane $e_p(r,z)$ on the color chart for the capped QD. A negative strain indicates that the distance between the atoms has decreased. Because Ge has a lattice parameter that is greater than that of Si, Table \ref{parametros}, the epitaxially grown Ge layers have a negative strain in plane $xy$ of the order of approximately $-4$\%. Just as the Si layer compresses the Ge QD, the Ge QD expands the Si layer, which is what we observed in the Si layers immediately above and below the Ge QDs where the strain reaches positive values higher than 0.5\%. 

Figure \ref{epzun} shows the spatial profile of $e_p(r,z)$ for the uncapped QD. The analysis follows the same trend: negative strain at the QD and positive one for the Si layer below the dot. However, the difference is that no Si layer is presented above the Ge QD, which results in a remarkable distinction: the strain in Ge is lower in module, i.e., while at the capped QD the strain is approximately $-3$\% at half height, it is approximately $-2$\% at the uncapped QD. We can understand this by considering that in the capped case, both the substrate and Si cover compress the Ge. In the uncapped case, only the substrate compresses the QD, and at the superior interface, the Ge is free; thus, its original lattice parameter tends to be reached more quickly. The behavior is different in the Si substrate, which is more strained under the uncapped dot. This condition can be understood as a reaction of the Ge dot, which in turn is less strained and possesses a more different lattice parameter that creates a larger strain on the Si layer. In the capped case, the Ge dot is more strained, which makes its lattice parameter more similar to the Si layer, creating a small strain on the Si layer. 

\begin{figure*}
\centering
\subfigure{a)
\includegraphics[width=3.0 in]{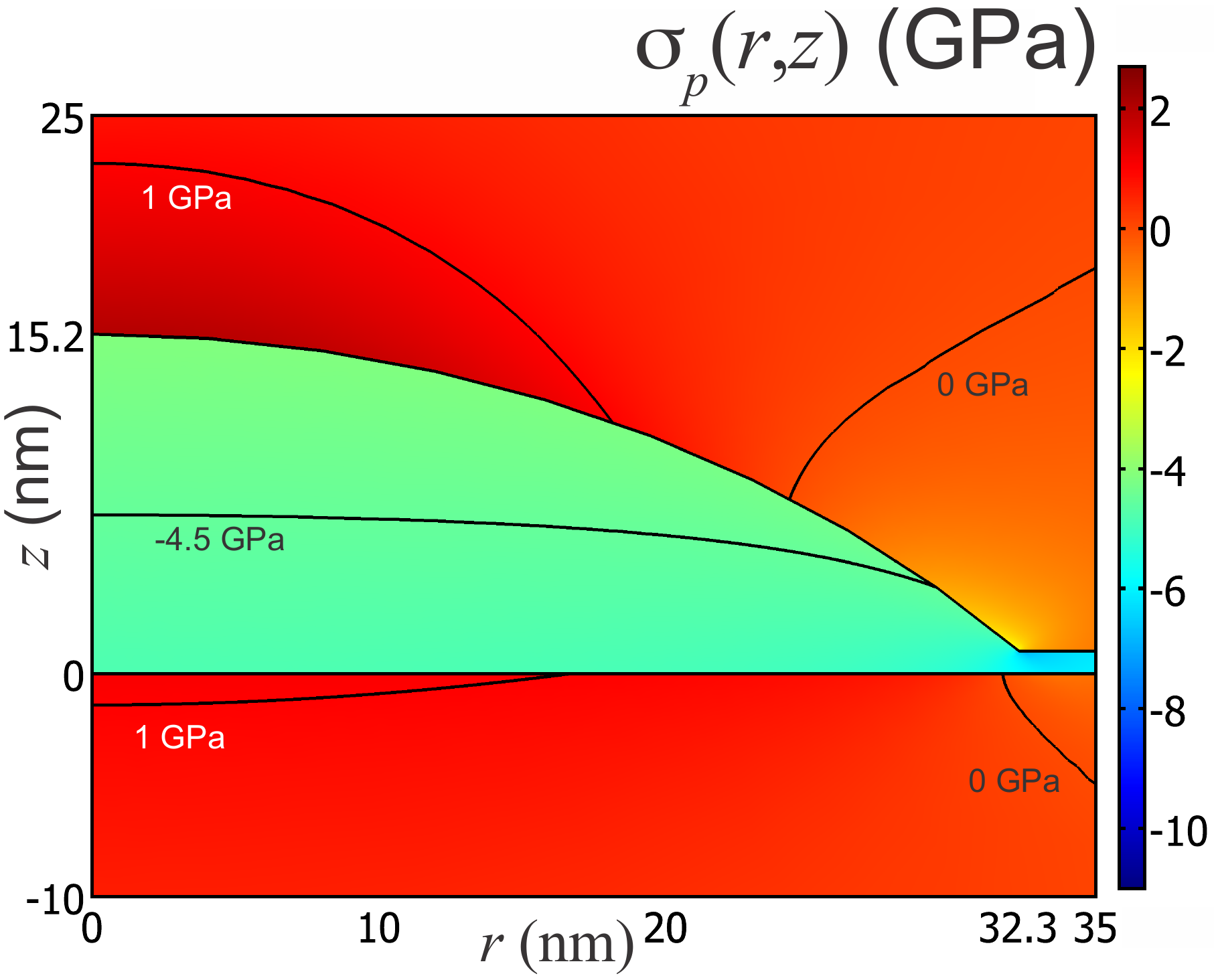}
\label{domo-sigmaplano-coberto}
}
\subfigure{b)
\includegraphics[width=3.0 in]{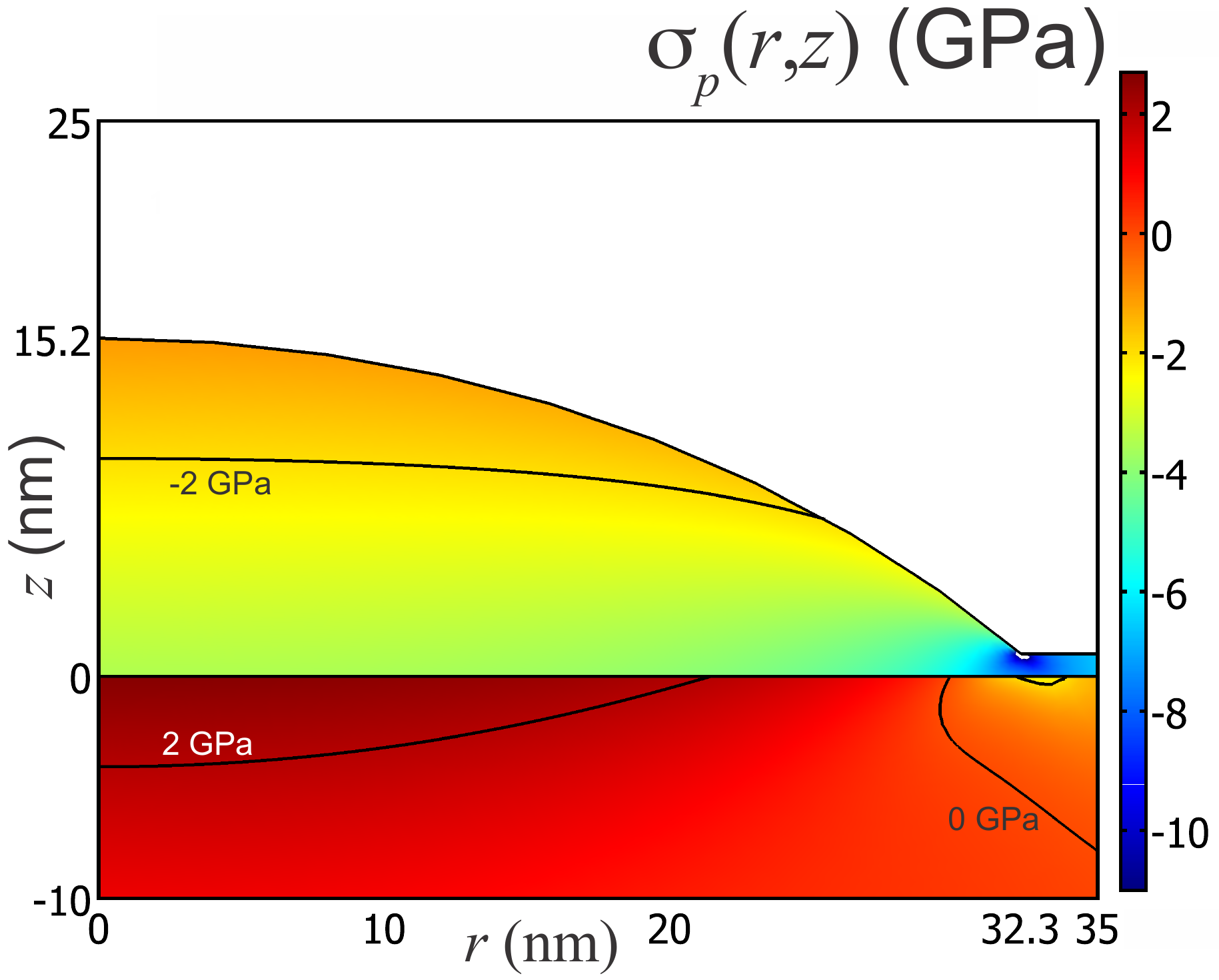}
\label{domo-sigmaplano-descoberto}
}
\caption{Spatial profile (in color) of the tension in the plane $\sigma_p (r,z)$ [Eq. (\ref{stress-plano})]. \subref{domo-sigmaplano-coberto} Capped QD. \subref{domo-sigmaplano-descoberto} Uncapped QD. The color scale bar is the same for both graphics.}
\end{figure*}

\begin{figure}
\centering
\includegraphics[width=3.0 in]{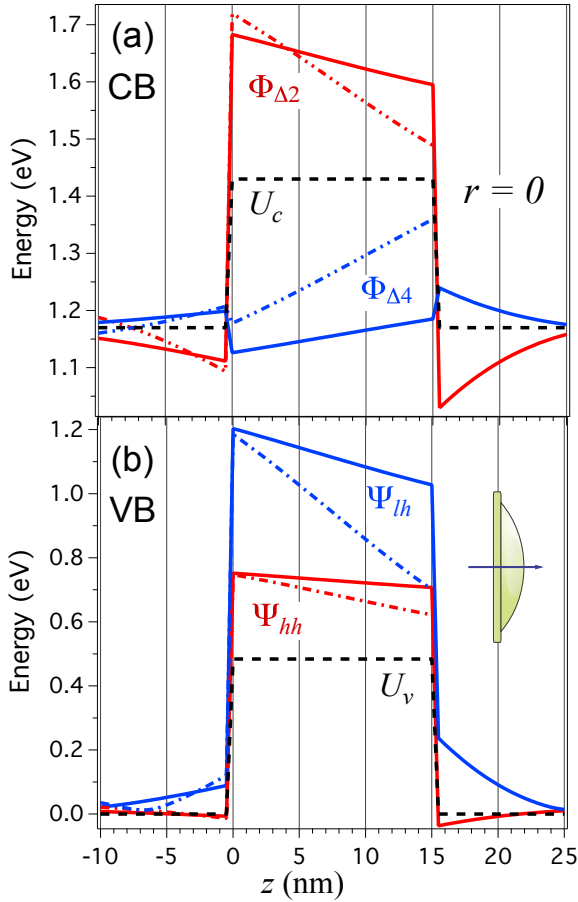} 
\caption{Band-edge alignment in the $z-$axis ($r=0$). Solid lines refer to the capped QD and dot-dashed lines represent the uncapped QD. The silicon substrate is the region $z<0$, the wetting layer is $0<z<1$, the QD is $1<z<15$.  (a) Conduction band (CB). Blue lines are $\Phi_{\Delta2}$ [Eq. (\ref{phidelta2})] and red lines are $\Phi_{\Delta4}$ [Eq. (\ref{phidelta4})]. Black dashed line is $U_c$. The cap layer is $z>15$. (b) Valence band (VB). Red lines are $\Psi_{hh}$ [Eq. (\ref{psihh})] and blue lines are $\Psi_{lh}$ [Eq. (\ref{psilh})]. Black dashed line is $U_v$. }
\label{bandas3}
\end{figure}

The $z-$component of the initial strain is proportional to the component in the plane but with an opposite sign. When the first layer of Ge is bound to the Si layer, compression exists in the plane of Ge, but no tension is present in the $z-$direction perpendicular to the plane. Therefore, the compression in the plane ($e_p<0$) implies an expansion in the $z-$direction ($e_z>0$) as if Ge tries to keep the volume of the unit cell costant. Figure \ref{epzca} shows the behavior of $e_z$ for the capped QD. One can see that a small compression ($e_z \sim -0.5$ \%) exists in the Si layer under the QD and a large expansion ($e_z \sim 2$ \%) exists inside the QD in response to the compression in the plane. In the same manner, the Si cap layer also exhibits compression ($\sim -1$ \%). In the uncapped case, shown in Fig. \ref{epzun}, no Si cap layer exists to exert an additional stress on the Ge QD, which makes the uncapped QD less strained than the capped QD, especially near its top surface. This condition makes the uncapped QD have a similar lattice parameter to its bulk value. Consequently, the uncapped QD exerts more tension on the Si substrate compared with the capped case.

The lattice parameter in the plane [Eq. (\ref{presplano})] is shown in Fig. \ref{domo-ap-coberto}. As expected, the lattice parameter in the Ge is smaller than its bulk value. At the top of the dot, $a_r$ reaches the maximum value for Ge, which is 5.5 \AA. In the same manner, in the Si above and below the dot, the parameter is larger than its bulk value (5.431 \AA). In both regions, $a_r$ reaches values greater than 5.45 \AA. Figure \ref{domo-ap-descoberto} shows the profile of $a_r$ for the uncapped dot. The lattice parameter in Ge attains higher values in the uncapped dot than that in the capped one. For instance, at the top of the uncapped dot, $a_r$ attains values exceeding 5.6 \AA. However, the maximum value of the capped dot (5.5 \AA) is only the minimum value of the uncapped dot. The Si also shows a difference. In the uncapped dot, the Si below the dot attains higher parameters of the order of 5.5 \AA.

Figure \ref{domo-sigmaplano-coberto} presents the calculated stress tension in the plane,
\begin{equation}
\sigma_p(r,z) = \frac{1}{2} (\sigma_r +\sigma_{\varphi}), \label{stress-plano} 
\end{equation}
defined as analogous to $e_p$. The components $\sigma_r$, $\sigma_{\varphi}$, and $\sigma_z$ are defined in Eq. (\ref{leidehookedetalhada}). Figure \ref{domo-sigmaplano-descoberto} shows $\sigma_p(r,z)$ for the uncapped dot. In the capped dot, the compressive tension in Ge is greater than that in the uncapped one. Conversely, in Si, the tension of traction (which tends to expand) is greater at the uncapped dot than that in the capped one. This is expected whereas in the capped dot, Ge undergoes more compression and thus manages to exert less traction (or tension of expansion) in Si at the capped QD.

Figures \ref{bandas3}{\color{blue}(a)} and \ref{bandas3}{\color{blue}(b)} show $U_c$ and $U_v$ (black dashed line) without strain contribution. In this situation, the electron is localized in the Si layers (substrate or cap layer) while the holes are confined in the QD. Figure \ref{bandas3}{\color{blue}(a)} shows the graphs of the electron energy levels $\Delta(2)$ and $\Delta(4)$ in the CB along the $z-$axis. For the capped QD, the minimum in the Si layer is in the $\Delta(2)$ band while inside the QD, the minimum is in the $\Delta(4)$ band. The split of these two bands are $\sim 83$ meV in the substrate and $\sim 480$ meV inside the QD. In the uncapped case, the minima in the substrate and inside the QD remain unchanged although the energy split inside the QD decreases to $\sim 336$ meV. Additionally, the band edge inside the uncapped QD has its minimum in the base of it, near the substrate. Other interesting feature is that the $\Delta(4)$ minimum inside the uncapped QD is $\Phi_{\Delta4}(z=2) - \Phi_{\Delta2}(z=-1)=83$ meV higher than the $\Delta(2)$ minimum in the substrate. In the capped case, this difference is only 14 meV. This means that the electron wave-function (which is localized in the Si layer) can penetrate much less inside the uncapped QD than in the capped one. A smaller penetration implies a lower superposition with the hole wave-function. This can be important to understand the different behavior of the band alignment (type I or type II) already studied for this system \cite{gomes2007,Larsson2006,Kondratenko2008}.

Figure \ref{bandas3}{\color{blue}(b)} shows the energy levels of the heavy and light hole in the VB. An interesting feature is the energy split of the bands,  $\Psi_{lh}-\Psi_{hh}$ at $z=7.5$ nm, which is 384 and 240 meV for the capped and uncapped cases, respectively. As the two holes are already confined within the QD, the effect of the strain does not change considerably their eigenstates. However, in experiments where an external strain is applied, the two bands ($hh$ and $lh$) shifts with different speeds. In the uncapped case, an energy split smaller than the above values can cause an anti-crossing of the two bands \cite{Gomes2004} with a smaller external strain.

\section{Conclusions}

We have calculated the spatial profile of mechanical strain in self-assembled Ge QDs grown on Si matrix. The Ge QD experiences a compressive biaxial strain in the $xy-$plane, which in turn creates a tensile strain on the $z-$plane. This effect is less intense in the uncapped QD because only one Si layer (substrate) compresses the Ge QD. Conversely, the Si substrate and cap layer experiences tensile strain on the $xy-$plane and compressive strain on the $z-$plane. In addition, the Si substrate is more strained in the uncapped case because the Ge uncapped dot is less strained. All these effects are reflected on the final lattice parameter and the stress tensor. We also calculated the band-edge alignment of the conduction and valence bands. The strain induced potentials suggest a smaller wave-function superposition in the uncapped QD when compared with the capped one.


\begin{thebibliography}{}


\bibitem{stangl} J. Stangl, V. Hol, G. Bauer, Review Modern Physics \href{http://link.aps.org/abstract/RMP/v76/p725}{\textbf{76} (2004) 725-783}.

\bibitem{boxberg2007}  F. Boxberg and J. Tulkki, Reports on Progess in Physics \href{http://iopscience.iop.org/0034-4885/70/8/R04/}{\textbf{70} (2007) 1425-1471}. 

\bibitem{jacobsen2006}  R. S. Jacobsen, K. N. Andersen, P. I. Borel, J. Fage-Pedersen, L. H. Frandsen, O. Hansen, Nature \href{http://www.nature.com/nature/journal/v441/n7090/full/nature04706.html}{\textbf{441} (2006) 199-202}.

\bibitem{ribeiro2007nanoletters} G. Medeiros-Ribeiro e R. S. Williams, NanoLetters \href{http://dx.doi.org/10.1021/nl062530k}{\textbf{7} (2007) 223}.

\bibitem{grundmann1995} M. Grundmann, O. Stier, and D. Bimberg, Physical Review B \href{http://link.aps.org/doi/10.1103/PhysRevB.52.11969}{\textbf{52} (1995) 11969-11981}.

\bibitem{andreev2000} A. D. Andreev and E. P. O'Reilly, Physical Review  B \href{http://link.aps.org/doi/10.1103/PhysRevB.62.15851}{\textbf{62} (2000) 15851-15870}.

\bibitem{jogai2001} B. Jogai, Journal of Applied Physics \href{http://link.aip.org/link/doi/10.1063/1.1379561}{\textbf{90} (2001) 699-704}.

\bibitem{cros2006} A. Cros, N. Garro, J. M. Llorens, A. García-Cristóbal, A. Cantarero, N. Gogneau, E. Monroy and B. Daudin, Physical Review B \href{http://link.aps.org/doi/10.1103/PhysRevB.73.115313}{\textbf{73} (2006) 115313}.

\bibitem{stier1999} O. Stier, M. Grundmann, D. Bimberg, Physical Review B  \href{http://link.aps.org/doi/10.1103/PhysRevB.59.5688}{\textbf{59} (1999) 5688-5700}.

\bibitem{he2004} L. He, G. Bester e A. Zunger, Physical Review B \href{http://link.aps.org/doi/10.1103/PhysRevB.70.235316}{\textbf{70} (2004) 235316}.

\bibitem{yakimov2010} A. I. Yakimov, A. A. Bloshkin and A. V. Dvurechenskii, Physical Review B \href{http://link.aps.org/doi/10.1103/PhysRevB.81.115434}{\textbf{81} (2010) 115434}.

\bibitem{brunner} K. Brunner, Reports on Progress in Physics \href{http://www.iop.org/EJ/abstract/0034-4885/65/1/202}{\textbf{65} (2002) 27}.

\bibitem{kurdi2006} M. El Kurdi, S. Sauvage, G. Fishman, and P. Boucaud, Physical Review B \href{http://link.aps.org/abstract/PRB/v73/e195327}{\textbf{73} (2006) 195327}.

\bibitem{gomes2007} P. F. Gomes, F. Iikawa, F. Cerdeira, M. Larsson, A. Elfving, G. V. Hansson, W.-X. Ni, and P.-O. Holtz, Applied Physics Letters \href{http://apl.aip.org/resource/1/applab/v91/i5/p051917_s1}{\textbf{91} (2007) 051917}.

\bibitem{ross1999} F. M. Ross, R. M. Tromp, M. C. Reuter, Science \href{http://www.sciencemag.org/cgi/content/full/286/5446/1931}{\textbf{286} (1999) 1931}.

\bibitem{gilbertoscience} G. Medeiros-Ribeiro, A. M. Bratkovski, T. I. Kamins, D. A. A. Ohlberg, R. S. Williams, Science \href{http://www.sciencemag.org/cgi/content/full/279/5349/353}{\textbf{279} (1998) 353}.

\bibitem{landau} L. D. Landau e E. M. Lifshitz, \textit{Theory of Elasticity}, 3rd Edition, Landau and Lifshitz Series Course of Theoretical Physics, volume 7, Elsevier, (1986).

\bibitem{sad2004} Martin H. Sad, \textit{Elasticity: Theory, Applications an Numerics}, Elsevier Academic Press (2004).

\bibitem{montoro2010} L. A. Montoro, G. Medeiros-Ribeiro e A. J. Ramirez, Journal of Physical Chemistry C \href{http://pubs.acs.org/doi/abs/10.1021/jp100187u}{\textbf{114} (2010) 12409-12415}.

\bibitem{kuo2008} M. K. Kuon T. R. Lin and K. B. Hong, Journal of Applied Physics \href{http://link.aip.org/link/?JAPIAU/103/073705/1}{\textbf{103} (2008) 073705}.


\bibitem{jalabert} D. Jalabert, J. Coraux, H. Renevier, B. Daudin, M.-H. Cho, K. B. Chung, D. W. Moon, J. M. Llorens, N. Garro, A. Cros, and A. García-Cristóbal, Physical Review B \href{http://link.aps.org/abstract/PRB/v72/e115301}{\textbf{72} (2005) 115301}.

\bibitem{Pryor1998}  C. Pryor, J. Kim, L. W. Wang, A. J. Williamson, and A. Zunger, Journal of Applied Physics \href{http://link.aip.org/link/doi/10.1063/1.366631}{\textbf{83} (1998) 2548.} 

\bibitem{Schmidt2000} O. G. Schmidt and K. Eberl, Y. Rau, Physical Review B \href{http://journals.aps.org/prb/pdf/10.1103/PhysRevB.62.16715#}{\textbf{62} (2000) 16715}.

\bibitem{Califano2002} M. Califano and P. Harrison, Journal of Applied Physics \href{http://dx.doi.org/10.1063/1.1410318}{\textbf{91} (2002) 389}.

\bibitem{paniago} R. Magalhes-Paniago, G. Medeiros-Ribeiro, A. Malachias, S. Kycia, T. I. Kamins, and R. Stan Williams, Physical Review B \href{http://link.aps.org/abstract/PRB/v66/e245312}{\textbf{66} (2002) 245312}.

\bibitem{Larsson2006} M. Larsson, A. Elfving, W.-X. Ni, G. V. Hansson, P. O. Holtz, Physical Review B \href{http://link.aps.org/abstract/PRB/v73/e195319}{\textbf{73} (2006) 195319}.

\bibitem{Kondratenko2008} S. V. Kondratenko, A. S. Nikolenko, O. V. Vakulenko, M. Ya Valakh, V. O. Yukhymchuk, A. V. Dvurechenskii and A. I. Nikiforov, Nanotechnology \href{http://www.iop.org/EJ/abstract/0957-4484/19/14/145703/}{\textbf{19} (2008) 145703}.

\bibitem{Gomes2004} P. F. Gomes, M. P. F. Godoy, M. K. K. Nakaema, F. Iikawa, T. E. Lamas, A. A. Quivy, and J. A. Brum, Physica Status Solidi (c) \href{10.1002/pssc.200304036}{\textbf{1} No. 3 (2004) 547.}





\end{thebibliography}
\end{document}